\documentclass{aa} 
\usepackage{graphicx,natbib,psfig,textcomp,fontenc,amssymb}  
\newcommand{\ltsima} {$\; \buildrel < \over \sim \;$} 
\newcommand{\gtsima} {$\; \buildrel > \over \sim \;$} 
\newcommand{\lta} {\lower.5ex\hbox{\ltsima}} 
\newcommand{\gta} {\lower.5ex\hbox{\gtsima}} 
\newcommand{\Ha} {H$\alpha$}
\newcommand{\Hb} {H$\beta$}

\newcommand{\ergsHz}{\ensuremath{{\rm erg}\,{\rm s}^{-1}\,{\rm Hz}^{-1}}}

\begin{document} 
\title{Radio and spectroscopic properties of miniature
  radio galaxies: revealing the bulk of the radio-loud AGN population.
\thanks{Based on observations made with the Italian Telescopio
    Nazionale Galileo operated on the island of La Palma by the Centro Galileo
    Galilei of INAF (Istituto Nazionale di Astrofisica) at the Spanish
    Observatorio del Roque del los Muchachos of the Instituto de Astrofisica
    de Canarias.}}
  
\titlerunning{Properties of miniature radio galaxies}

\authorrunning{R.~D. Baldi \& A. Capetti}
  
\author{Ranieri D. Baldi
\inst{1}
\and  Alessandro Capetti \inst{2}} 
\offprints{R.D. Baldi}  
\institute{
Universit\'a di Torino, via P. Giuria 1, 10125 Torino, Italy\\
\email{baldi@oato.inaf.it}
\and 
INAF - Osservatorio Astronomico di Torino, Strada
  Osservatorio 20, I-10025 Pino Torinese, Italy\\
\email{capetti@oato.inaf.it}}

\date{}  
   
\abstract{We explore radio and spectroscopic properties of a sample of 14
  miniature radio galaxies, i.e. early-type core galaxies hosting radio-loud
  AGN of extremely low radio power, 10$^{27-29}$ erg s$^{-1}$ Hz$^{-1}$ at 1.4
  GHz.

  Miniature radio galaxies smoothly extend the relationships found for the
  more powerful FR~I radio galaxies between emission line, optical and radio
  nuclear luminosities to lower levels.  However, they have a deficit of a
  factor of $\sim$100 in extended radio emission with respect to that of the
  classical example of 3CR/FR~I. This is not due to their low luminosity,
  since we found radio galaxies of higher radio core power, similar to those
  of 3CR/FR~I, showing the same behavior, i.e. lacking significant extended
  radio emission.  Such sources form the bulk of the population of radio-loud
  AGN in the Sloan Digital Sky Survey. At a given level of nuclear emission,
  one can find radio sources with an extremely wide range, a factor of
  $\gtrsim$100, of radio power.

  We argue that the prevalence of sources with luminous extended radio
  structures in flux limited samples is due to a selection bias, since the
  inclusion of such objects is highly favored. The most studied catalogues of
  radio galaxies are thus composed by the minority of radio-loud AGN that meet
  the physical conditions required to form extended radio sources, while the
  bulk of the population is virtually unexplored.

\keywords{Galaxies: active -- Galaxies: elliptical
    and lenticular, cD -- Galaxies: nuclei -- Galaxies: jets -- Galaxies: evolution}}

\maketitle
  
\section{Introduction}

The presence of a radio source represents the most common manifestation of
nuclear activity in early type galaxies. These sources are associated with a
fraction as high as 40\% of all bright elliptical and lenticular galaxies
(e.g. \citealt{sadler89,auriemma77}).  Due to the steepness of their
luminosity function, most radio sources are of low luminosity. The interest in
the properties of these objects lies in the fact that they sample an
essentially unexplored regime for radio loud AGN (hereafter RLAGN) and they
effectively close the gap between active and quiescent galaxies.

\citet{balmaverde06b} considered a sample of low luminosity radio sources
($\sim 10^{26-29}$ erg s$^{-1}$ Hz$^{-1}$ at 5 GHz) well suited for such a
study. They are hosted by early-type galaxies selected on the basis of the
presence of a shallow core in their host surface brightness profiles, defined
as ``core'' galaxies (hereafter CoreG).  They used HST and Chandra data to
isolate their optical and X-ray nuclear emission, showing that CoreG
invariably host radio-loud nuclei, with an average radio loudness parameter of
$R = L_{5\rm {GHz}} / L_{\rm B}$ $\sim$ 4000, similar to the value measured
for FR~I radio galaxies. Their optical and X-ray nuclear luminosities
correlate with the radio-core power, smoothly extending the analogous
correlations found for FR~I low luminosity radio galaxies
\citep{chiaberge:ccc,balmaverde06a} toward even lower power, by a factor of
$\sim$1000, covering a combined range of 6 orders of magnitude.

The similarities between CoreG and FR~I include the distributions of black
hole masses, host galaxy luminosities, and also the properties of the
surface brightness profiles \citep{deruiter05}. This indicates that they are
drawn from the same population of early-type galaxies.

The level of activity in these sources is closely related to the accretion
rate of hot gas derived analyzing Chandra images.  \citet{balmaverde08}
considered a sample of 44 galaxies (CoreG and FR~I) and found that the
accretion power correlates linearly with the jet power. These results
strengthen and extend the validity of the results obtained by \citet{allen06}
and \citet{hardcastle07}, indicating that hot gas accretion is the dominant
process in powering FR~I radio galaxies across their full range of
radio-luminosity. CoreG follow the same relationship, scaled to
correspondingly lower accretion power.

This result, combined with the analogy of the nuclear properties, lead
\citet{balmaverde06b} to the conclusion that CoreG can be effectively
considered as miniature radio galaxies.  The exploration of the properties of
these sources offers the opportunity to probe the physical properties of RLAGN
at widely different levels of activity with respect to classical FR~I
radio galaxies. The aim of this paper is to expand the study of
\citet{balmaverde06b} by including i) an optical spectroscopic
characterization of CoreG (derived by measuring ratios of emission lines) and
ii) a detailed analysis of their radio properties (considering both
morphologies and spectra), in order to perform a comparison with radio
galaxies of higher power.

The main finding of this study is that while from the nuclear point of view
CoreG are simply scaled down version of FR~I they have a deficit of a factor
of $\sim$100 in extended radio emission with respect to that of classical
FR~I. We will show that, rather surprisingly, this is a general property, not
limited to these faint radio sources, and that the vast majority of the
population of RLAGN does not show the characteristic very luminous extended
radio structures.

The paper is organized as follows. In Sect.  \ref{sample} we present the
sample selection. In Sect. \ref{radiop} and \ref{spectrop} we analyze the
radio and optical spectroscopic properties of the CoreG. This leads to a
multiwavelength view of their nuclei (Sect. \ref{multip}) that is compared
with those of the more powerful 3CR/FR~I. What emerges is that CoreG have a
substantial deficit of extended radio emission. In Sect.  \ref{bulk} we show,
considering different samples of radio sources, that radio galaxies with
feeble extended structures represent the bulk of radio-loud AGN population. In
Sect.  \ref{discussion} we discuss the consequences of this result on the
general properties of radio galaxies and on our understanding of their nuclear
activity. In Sect.  \ref{summary} we summarize our findings and present our
conclusions. In two Appendices we describe the details of the optical
spectroscopic observations and explore the multiphase interstellar medium of
CoreG.

\section{Sample selection}
\label{sample}

\citet{early1} considered two samples (located in the Northern and Southern
hemispheres) of luminous ($B \leq$ 14), nearby ($V_{\rm rec} <$ 3000)
early-type galaxies for which extensive multiwavelength data from VLA, HST,
and Chandra observations are available.  They selected the galaxies detected
in the VLA surveys of \citet{wrobel91b} and \citet{sadler89} at a flux limit
of $\sim$1 mJy at 5 GHz, and used archival HST observations to study their
surface brightness profiles.  \citet{balmaverde06b} focused on the 29 ``core''
galaxies, characterized by an flat inner logarithmic slope ($\gamma$
$\leqslant$ 0.3). As discussed in the Introduction they represent miniature
versions of radio galaxies.

We extracted the Northern part of this sample of core galaxies (17 objects)
for our observing programs, by leaving aside only 
the 3 powerful radio galaxies part of the 3CR
sample, namely UGC~7360 (3C~270), UGC~7494 (3C~272.1, M~84), and UGC~7654
(3C~274, M~87). The final sample consists of 14 objects.  

\section{Radio properties of core galaxies}
\label{radiop}

In order to analyze the radio properties of the CoreG, we first search in the
literature for their radio maps, looking for extended emission.  Twin jets are
seen in the VLA images of UGC~7629, UGC~7878, UGC~7898, and UGC~8745 (the
object descriptions with relative references and measured fluxes are given in
Table~\ref{tab1}), while all other sources are unresolved. Nonetheless, higher
resolution data obtained with the VLBA \citep{nagar05} detected sub-parsec
scale jets for UGC~7386, UGC~7760, UGC~7797, UGC~9706.  UGC~968, UGC~7203, and
UGC~9723 are still unresolved even on the VLBA scale. No information on the
parsec scale is available for UGC~5902, UGC~6297, and UGC~9655.

In general, the extended radio emission of these sources is dominated by the
presence of jets, similar in morphology to those seen in FR~I galaxies, but on
smaller scales. The CoreG found to be extended in VLA images have sizes of
11-22 kpc, to be compared with the FR~I part of 3C catalogue
\citep[see][]{chiaberge:ccc} whose sizes range from 8 to 400 kpc with a median
of 100 kpc.

\begin{figure*}
\includegraphics[scale=0.45]{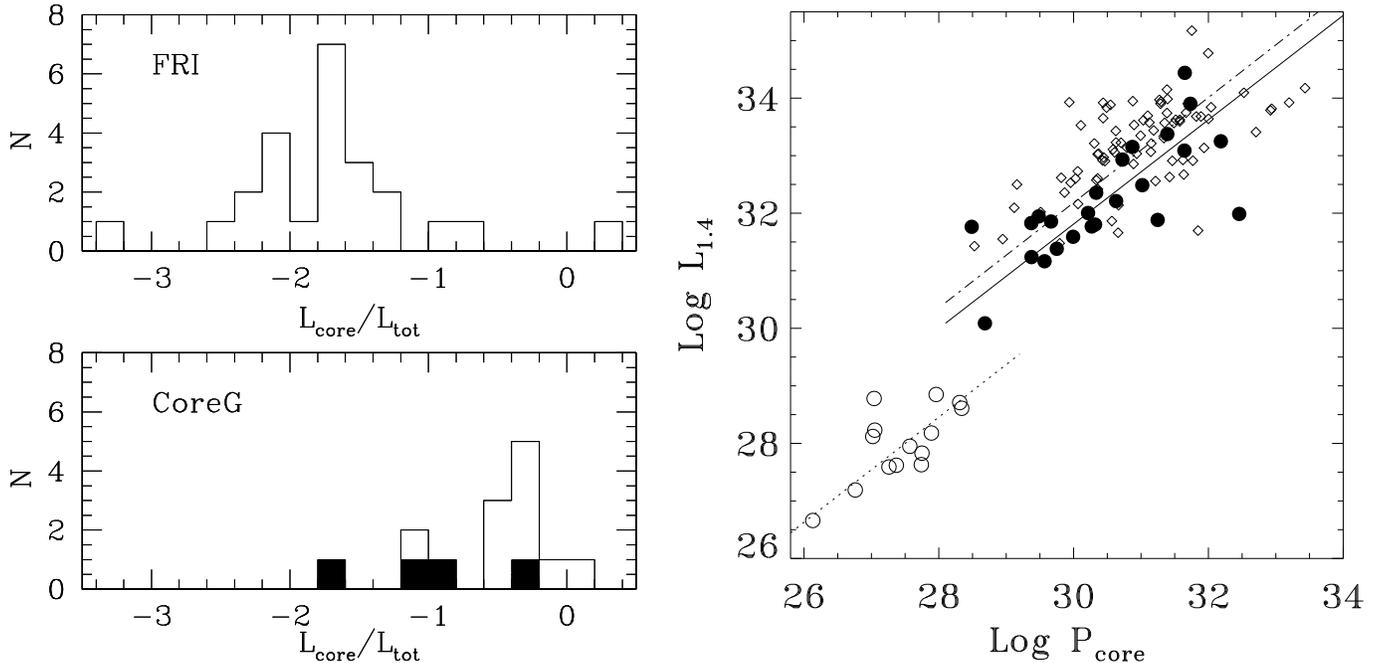}
\includegraphics[scale=0.725]{13021f1b.epsi}
\caption{Left: histograms of 
core dominance $P_{\rm c} = P_{\rm core} / L_{1.4}$ for the
  3C/FR~I and CoreG samples. Right: Core radio power at 5 GHz vs. total
  emission at 1.4 GHz for CoreG (empty circles), 3CR radio galaxies
  (diamonds) where those with FR~I
  morphology are indicated as filled circles. Lines indicate the 
best linear fit between the two quantities considered, dotted for CoreG,
dot-dashed for 3CR, solid for 3CR/FR~I.}
\label{cd}
\end{figure*}

Using the total 1.4 GHz ($L_{1.4}$) and the 5 GHz core fluxes ($P_{\rm core}$)
derived for several objects from our own A array VLA observations at 5 GHz
(Baldi et al. 2009 in preparation), see Table \ref{tab2}, we estimated the
core dominance of the sources of the sample, i.e. $P_c$ = $P_{\rm
core}/L_{1.4}$.  Most objects have a very high core dominance (Log $P_c$ = -1
and 0) compared to that measured in 3C/FR~I objects (see Fig.~\ref{cd}). Three
of the four objects with large scale jets (UGC~7629, UGC~7878, and UGC~8745)
have, together with UGC~9655, the lowest core dominance, reaching the tail of
the high core dominance of FR~I.

To explore in more detail the core dominance of CoreG we compare in the right
panel of Fig.~\ref{cd} their core and total radio luminosities with those of
all 3CR radio galaxies up to a redshift of z=0.3. The core dominance in CoreG
is a factor of $\sim$ 100 higher than measured for the 3CR sample, or a factor
of $\sim$40 considering only the sample of 3CR/FR~I selected by
\citet{chiaberge:ccc}\footnote{From the original sample of
  \citet{chiaberge:ccc} we excluded i) 3C~277.3, 3C~305, and 3C~433 since from
  an accurate inspection of its radio maps we revised its radio-morphological
  classification from FR~I to FR~II; ii) 3C~28, 3C~314.1, and 3C~348 because
  they belong to the spectroscopic class of Extremely Low Excitation Galaxies
  (ELEG) \citep{buttiglione09b}; iii) and 3C~386 because of the chance
  superimposition of a star on the galaxy's nucleus
  \citep{lynds71,buttiglione09a}.}.

\begin{table*}
\begin{center}
\caption{Radio properties of the core galaxies.}
\begin{tabular}{c| c c c c c | c}
\hline
Name& $F_{\rm core}$ & $F_{\rm 74 MHz}$ & $F_{\rm 1.4 GHz}$ & $F_{\rm 5 GHz}$ & $P_c$ & radio structure\\
    &  mJy           &  mJy             & mJy               & mJy             &       &                \\ 
\hline
UGC~5902  & 0.73$^{a}$  &        & 2.4   & 0.7   &  0.30   & VLA unresolved$^{e}$ \\          
UGC~7203  & 4.8$^{b}$   &        & 5.6   & 4.5   &  0.86   & VLBI unresolved$^{m}$ \\           
UGC~7629  & 4.8$^{a}$   &        & 256   & 95	 &  0.02   & twin kpc jets$^{j}$\\            
UGC~7760  & 131.7$^{b}$ &        & 100   & 121   &  1.30   & twin pc jets$^{e}$ \\              
UGC~7898  & 12.5$^{a}$  & 449    & 29.1  & 24	 &  0.43   & twin kpc jets$^{k}$ \\           
UGC~8745  & 10.5$^{a}$  & 470    & 78.4  & 20	 &  0.13   & twin kpc jets$^{l}$\\            
UGC~9655  & 1.2$^{a}$   &        & 14.8  & 2.1   &  0.08   & VLA unresolved$^{h}$\\           
\hline   	      	     	     	    	                                                            
UGC~0968  & 1.5$^{c}$   &        & 3.1   & 1.4   &  0.48   & VLBI unresolved$^{i}$ \\           
UGC~6297  & 2.7$^{a}$   &        & 6.9   & 2.6   &  0.39   & VLA unresolved$^{h}$ \\          
UGC~7386  & 159.8$^{b}$ & 343    & 385   & 351   &  0.41   & pc jet$^{f,g}$ \\                  
UGC~7797  & 20.4$^{a}$  &        & 36.8  & 21	 &  0.55   & twin pc jets$^{h}$\\               
UGC~7878  & 5.3$^{a}$   & 692    & 77.8  & 45	 &  0.07   & twin kpc jets$^{k}$\\            
UGC~9706  & 6.4$^{d}$   &        & 21.0  & 5.3   &  0.30   & pc jet$^{i}$ \\                    
UGC~9723  & 12.7$^{b}$  &        & 21.8  & 7.4   &  0.58   & VLBI unresolved$^{n}$\\
\hline
\end{tabular}
\label{tab2}

\medskip

Column description: (1) name; (2) core flux at 5 GHz in mJy. References: (a)
  Baldi et al. 2009, in preparation; (b) \citet{nagar01}; (c) \citet{nagar05}
  (VLBA); (d) \citet{filho04} at 8 GHz; ; (3) total flux at 74 MHz in mJy; (4)
  total flux at 1.4 GHz from \citet{condon02} in mJy; (5) total flux at 5 GHz
  from \citet{wrobel91b} and \citet{wrobel91a} in mJy; (6) core dominance, (7)
  radio morphology.  References: (e) \citet{nagar02}; (f) \citet{jones84}; (g)
  \citet{falcke00}; (h) \citet{nagar05}; (i) \citet{filho04}; (j)
  \citet{ho01c}; (k) \citet{stanger86}; (l) \citet{hummel84}; (m)
  \citet{anderson04}; (n) \citet{anderson05}.
\end{center}
\end{table*}

\begin{figure*}
\includegraphics[scale=0.45]{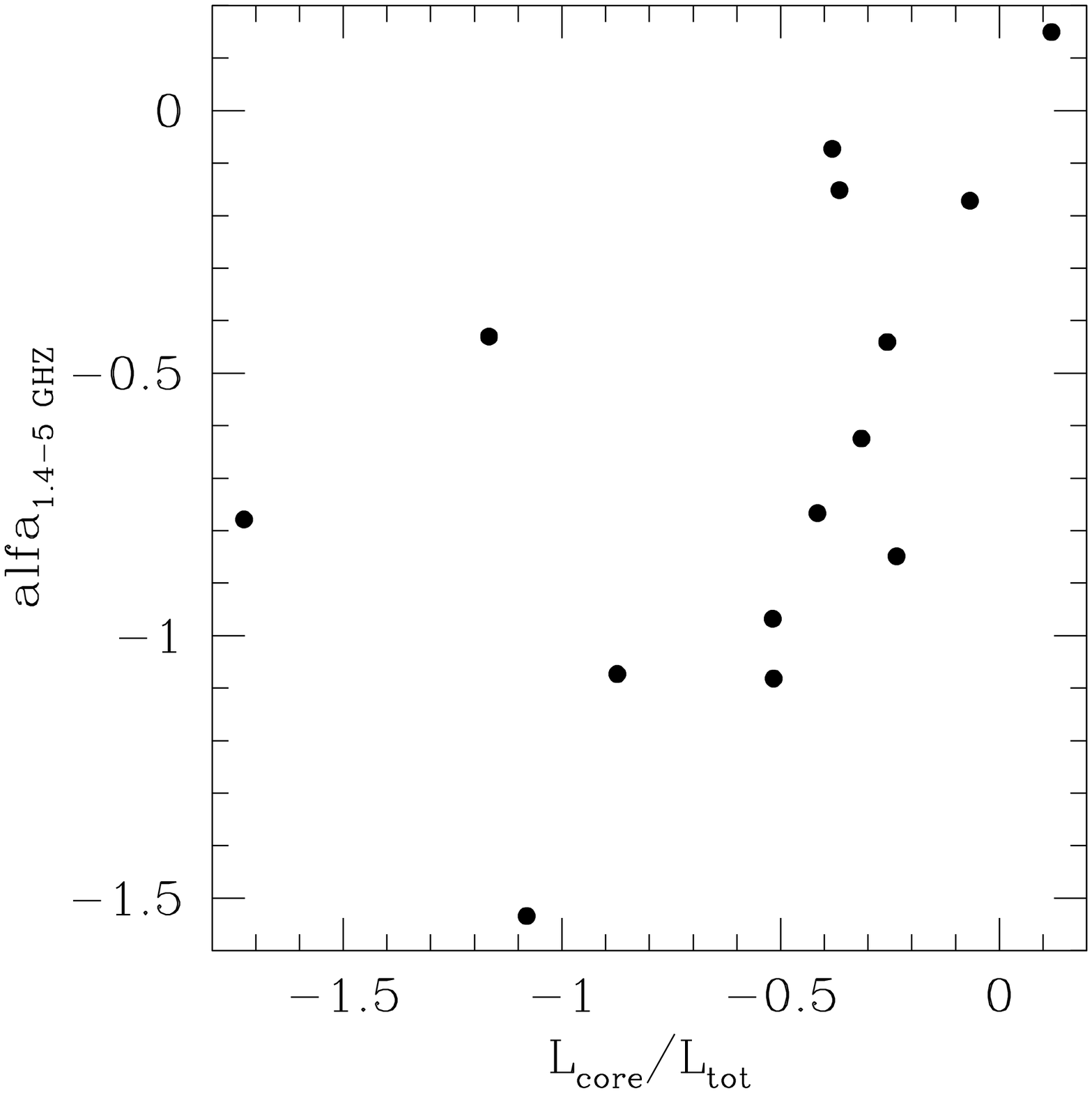}
\includegraphics[scale=0.45]{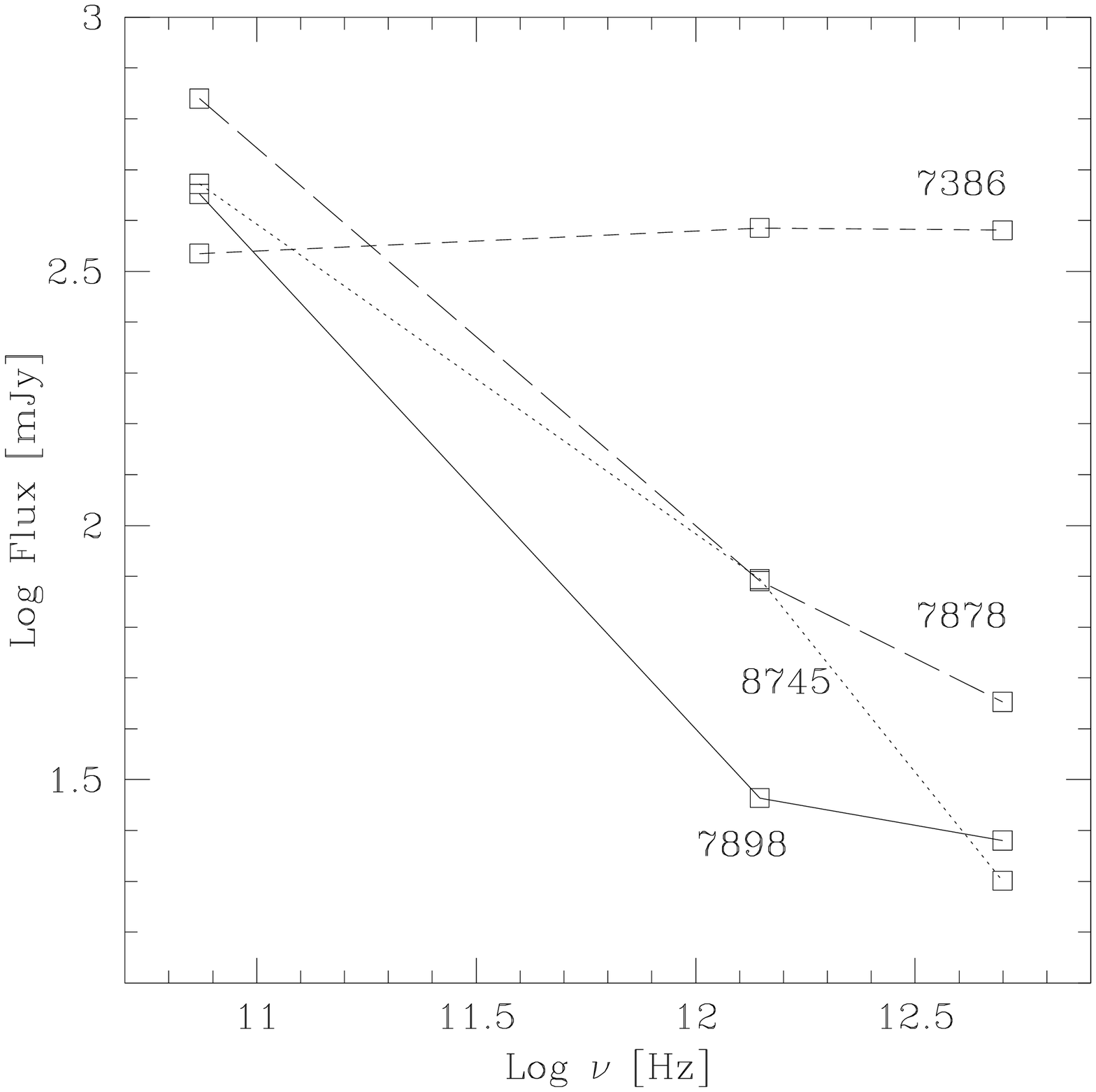}
\caption{Left: core dominance vs. spectral index between
  1.4 and 5 GHz for the CoreG sample. Right: 3-point radio spectra of the 4
  CoreG detected by the VLSS at 74 MHz.}
\label{cd2}
\end{figure*}

Let us now focus on the radio spectral properties of the CoreG. We
calculated the spectral index for the total emission between 1.4 and 5 GHz.
Not surprisingly, the spectral index has a strong dependence on the core
dominance (Fig.~\ref{cd2}, left panel). Objects in which the flat spectrum
core emission accounts for a large fraction of the total radio luminosity have
lower spectral indices.

Four objects have been detected by the VLA Low-frequency Sky Survey at 74 MHZ
\citep{cohen07}, namely UGC~7898, UGC07878, UGC~8745, and UGC~7386 (at a flux
exceeding the 3$\sigma$ limit of 0.3 Jy) and for these sources it is possible
to build a three-point radio spectrum (see Fig. \ref{cd2}, right panel).  The
first 3 have rather steep spectra, with $\alpha_{74,1400} \sim 0.6 - 0.9$.
These values are well within the range of the low frequency radio spectral
indices measured by \citet{kellermann69} for the 3CRR radio galaxies,
$\alpha_{38,750} \sim 0.75 \pm - 0.15$. In UGC
7898 the spectrum is clearly convex, indicative of a transition from an
optically thick (core) emission to the optically thin contribution of the
extended structure at low frequencies. Instead, UGC~7386 still has a flat
spectral index (-0.04), symptomatic of an extreme core dominance even at 74
MHz.

Therefore, although CoreG often show extended emission, often in the form of
well defined jets, it is less extended and less dominant (with respect to the
nuclear radio component) than in powerful radio sources. This effect cannot be
ascribed in general to a different spectral behavior of CoreG with respect to
FR~I, as they show similar low frequency spectral indices.  However, we cannot
distinguish between the two alternative explanations for these results, namely
that they are caused by Doppler boosting of the radio core or by a genuine
deficit of extended radio emission.

\section{Spectroscopic properties of Core Galaxies}
\label{spectrop}

The spectroscopic diagnostic diagrams, planes formed by pairs of emission line
ratios, can be used to assess the nature of the nuclear emission, e.g. 
separating active nuclei from star forming galaxies 
\citep[e.g.][]{baldwin81}. 
More recently, \citet{kewley06} selected a
sample of $\sim 85000$ emission line galaxies from the SDSS, finding that
Seyferts and LINERs \citep{heckman80} 
form separate branches in the diagnostic diagrams.

The spectroscopic survey of \citet{ho97} covers 13 out of 14 of the selected
CoreG. However, the spectra of several galaxies are of insufficient quality to
extract accurate line measurements and to obtain a robust spectral
classification. We then re-observed 7 of these sources at the Telescopio
Nazionale Galileo (TNG). The details of the observations are given in Appendix
A. We obtained data at a resolution of $R\sim$800 over the spectral range of
4650-6800 \AA\ and extracted the spectra over a 2\arcsec$\times$2\arcsec\
nuclear region.  We subtracted the contribution of the stellar emission using
as a template two off-nuclear regions, 2\arcsec\ wide, flanking the nuclear
aperture.

In all galaxies we were able to detect the \Ha, \Hb, and [N II] lines and with
only two exception (UGC~7898, and UGC~9655) we also measured the [O III] line.
Conversely, only one spectrum (UGC~7203) yields an estimate of [O I].  In
Table~\ref{tabriga} we report the line measurements for these 7 objects as
well as those presented by \citet{ho97} for the remainder of the sample.

\begin{figure*}
\centerline{
\includegraphics[scale=0.8,angle=90]{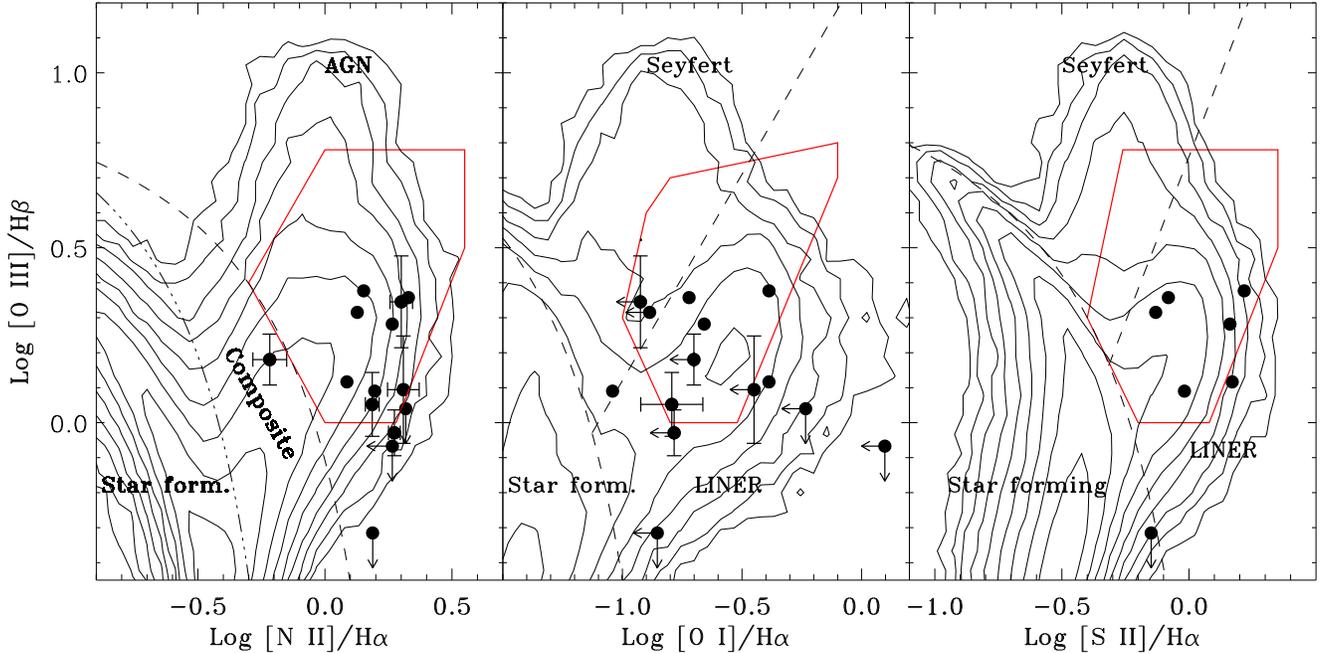}
}
\caption{Spectroscopic diagnostic diagrams: in left panel,
log([O~III]/H$\beta$) versus log([N~II]/H$\alpha$), in middle panel
log([O~III]/H$\beta$) versus log([O~I]/H$\alpha$), and in right panel,
log([O~III]/H$\beta$) versus log([S~II]/H$\alpha$). The dashed separation
lines among star forming galaxies, LINERs, and Seyferts is derived from
\citet{kewley06}, while the red box is the region covered by LEG in the 3CR
sample \citep{buttiglione09b}. In the left panel the region between the
dot-dashed and dashed lines is populated by composite galaxies, whose spectra
contain significant contributions from both AGN and star formation.}
\label{diagn}
\end{figure*}

In general, the objects of our sample occupy a region in the first diagnostic
diagram (see Fig.~\ref{diagn}) indicative of an AGN origin of their emission
lines. There are however a few possible exceptions: UGC~7898 only has
measurements for the \Ha\ line, but the limits of the diagnostic ratios are
consistent with it being an AGN. Furthermore, UGC~5902 falls, in the first
diagram, in the region of composite galaxies defined by \citet{kewley06},
possibly the location of objects intermediate between star-bursts and AGN.

The second and third diagrams are less well defined, due to the smaller number
of detections. In particular, since the wavelength coverage of our spectra
does not include the [S~II] lines, it is possible to include in the diagnostic
diagram of log([O~III]/H$\beta$) vs. log([S~II]/H$\alpha$) only the 7 sources
from \citet{ho97}. Nonetheless, the location of the points is generally
closely consistent with the region of LINERs according to the classification
of \citet{kewley06}.

However, the AGN considered by \citet{kewley06} are mostly radio-quiet, while
CoreG are radio-loud. Therefore, it is also instructive to perform a
comparison with the results of \citet{buttiglione09b} on the spectroscopic
properties of radio galaxies from the 3CR sample. Following previous studies,
(e.g. \citealt{hine79,laing94,jackson97}), they separated low and high
excitation galaxies (LEG and HEG respectively) on the basis of the narrow
emission line ratios. In particular, the location of LEG in the diagnostic
diagrams is represented by the red polygon in Fig.~\ref{diagn}.  Most of the
CoreG are included in this area, generally populating its lower half, but also
extending toward lower [O~III]/H$\beta$ ratios.

\citet{buttiglione09b} found that the separation between HEG and LEG
respectively is similar to that found by \citet{kewley06} for LINERs and
Seyferts among the SDSS sources. However, they found a significant number of
LEG located above the line marking the transition between LINERs and
Seyferts. The location of LEG shows an upward scatter with respect to the
`finger' of highest LINER density by $\sim$0.2 dex in the [O III]/\Hb\
ratio. Their data were not sufficient to conclude whether this was due to a
genuine difference between the (mostly) radio-quiet AGN of the SDSS and the
RLAGN of the 3CR sample, or simply to a luminosity difference. In fact, there
is a substantial mismatch in luminosity between the 3CR and the SDSS sources,
the former being brighter on average by a factor of $\sim$ 30.

The miniature radio galaxies
considered here are instead well matched with the SDSS AGN. The majority of
LINERs has luminosities in the range $10^{38}<L_{\rm [O~III]}<10^{39}$ erg
s$^{-1}$, while the sample of miniature radio galaxies discussed here has a
median luminosity of $L_{\rm [O~III]} \sim10^{38.3}$ erg s$^{-1}$. Indeed there
is a close overlap between the two classes in the spectroscopic diagnostic
diagrams.  This indicates that the line ratios are driven mostly by the
luminosity of the AGN rather than by its radio-loudness.

The main result derived from this analysis is that the observed line emission
generally has an AGN origin, and it is not associated with star formation.
This enables us to include the emission line luminosities in our analysis of
AGN properties.

\section{A multi-wavelength view of core galaxies}
\label{multip}
We collect multi-wavelength information for our sample of core galaxies in
order to compare the emission in the different bands and to contrast it with
the properties of the more powerful FR~I radio galaxies. All data used for
this analysis are reported in Table \ref{tab1}.

\begin{table}
\caption{Multiwavelength properties of the core galaxies.}
\begin{center}
\begin{tabular}{c| c c c c }
\hline
Name& $L_{\rm o}$& $P_{\rm core}$& $L_{1.4 GHz}$ & $L_{[O~III]}$ \\
\hline	                              
UGC~5902  &$<$39.10& 35.83&    35.81   &    37.96  \\
UGC~7203  & 39.85  & 37.45&    36.98   &    38.08  \\
UGC~7629  & 38.79  & 36.74&    37.93   &    37.40  \\
UGC~7760  & 38.73  & 37.44&    36.78   &    37.28  \\
UGC~7898  &$<$39.13& 37.27&    37.10   & $<$37.34  \\
UGC~8745  &  $-$   & 37.66&    38.00   &    38.41  \\
UGC~9655  &  $-$   & 36.72&    37.27   & $<$38.04  \\
\hline	 	   		            
UGC~0968  &$<$39.77& 36.96&    36.74   & $<$37.75  \\
UGC~6297  &$<$39.06& 36.46&    36.34   &    38.21  \\
UGC~7386  & 38.76  & 38.01&    37.86   &    38.95  \\
UGC~7797  &$<$40.19& 38.04&    37.76   &    38.79  \\
UGC~7878  & 39.07  & 36.75&    37.38   &    37.89  \\   
UGC~9706  & 39.25  & 37.59&    37.33   &    38.18  \\ 
UGC~9723  & $-$    & 37.07&    36.77   &    37.33  \\            
\hline
\end{tabular}

\medskip

Column description: (1) name; (2) optical nuclear luminosity [erg
  s$^{-1}$] from \citet{balmaverde06b}; (3) radio core power [erg s$^{-1}$] at
  5 GHz and (4) total radio luminosity at 1.4 GHz [erg s$^{-1}$] estimated from
the fluxes given in Tab. \ref{tab1}; (5) [O III] emission line luminosity
[erg s$^{-1}$].
\label{tab1}
\end{center}
\end{table}

As discussed, \citet{balmaverde06b} found that the optical and X-ray nuclear
luminosities correlate with the radio-core power, smoothly extending the
analogous correlations already found for 3CR/FR~I radio galaxies
\citep{chiaberge:ccc,hardcastle00,balmaverde06b} toward even lower power, by a
factor of $\sim 1000$, covering a combined range of 6 orders of magnitude.
Apparently CoreG are simply scaled down version of 3CR/FR~I from the point of
view of their nuclear emission. This consideration also applies to the level
of accretion of hot gas estimated by \citet{balmaverde08} in a sample of
radio galaxies that includes both CoreG and 3CR/FR~I. From the
spectroscopic observations we also estimated that the warm ionized gas
represents a small fraction ($10 ^{-4}$ - 7$\times 10 ^{-3}$) of the hot gas
component (see Appendix B). This strengthens a posteriori the validity of the
estimates of accretion rates derived considering only the high temperature
gas.

\begin{figure*}
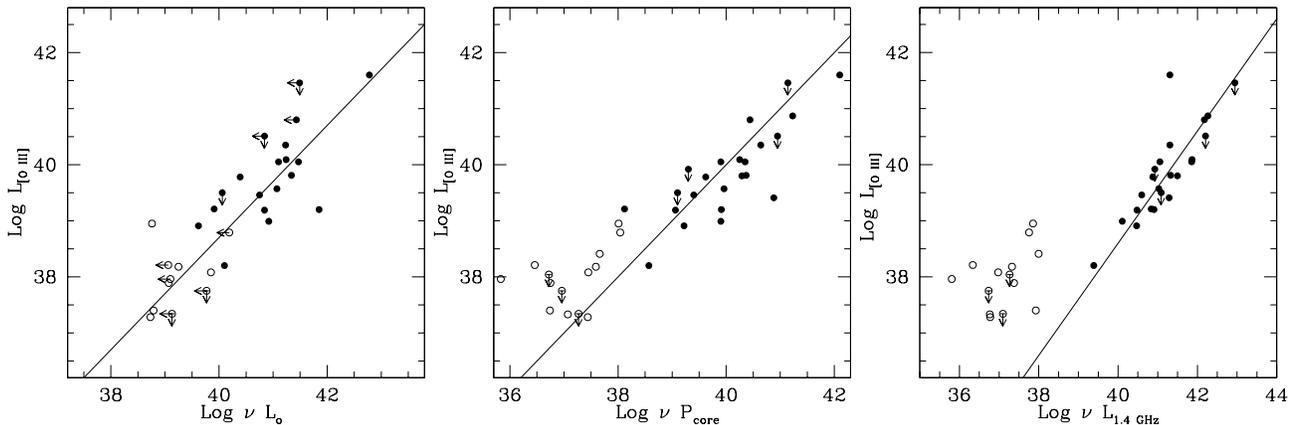

\centerline{\includegraphics[scale=0.29]{13021f4a.epsi}
\includegraphics[scale=0.29]{13021f4b.epsi}
\includegraphics[scale=0.29]{13021f4c.epsi}}
\caption{[O III] line luminosity [erg s$^{-1}$] compared to the luminosities 
  of (left) optical nucleus, (center) radio nucleus, and (right) total radio
  power at 1.4 GHz for Core galaxies (empty points) and 3CR/FR~I (filled
  points). The solid lines are parallel to the planes bisectrix and mark the
  loci of constant ratio between the two quantities.}
\label{nucleiriga}
\end{figure*}

The spectroscopic analysis of the CoreG rules out the star formation origin of
the nuclear line emission. Therefore we can include the emission line
luminosities in our analysis of the AGN properties. In Fig.~\ref{nucleiriga}
we show the [O~III] emission line luminosities of the miniature radio galaxies
with respect to the power of their radio core, optical nucleus, and total
radio emission.  As a comparison, we also present the same quantities for the
sample of 3CR/FR~I radio galaxies, using the line luminosities measured by
\citet{buttiglione09a}.

The radio and the optical nuclear luminosities are well correlated with the
emission line luminosities for both CoreG and 3CR/FR~I, with CoreG extending
these trends toward luminosities up to $\sim$100 lower with respect to
3CR/FR~I radio galaxies.  Conversely CoreG do not follow the relationship
between total radio power and emission line defined by 3CR/FR~I, showing an
excess in line or a deficit in radio emission typically by a factor of $\sim$
100.

In Sect. \ref{radiop} we found a ratio between total and core radio emission
approximately two orders of magnitude lower in CoreG than in 3CR/FR~I radio
galaxies. From the radio data alone the possibility that this is due to a
Doppler boosting enhancement of the radio core emission remains viable.
However, the inclusion of the emission line (a quantity independent of
orientation) leads us to the conclusion that what we are seeing in CoreG is a
genuine deficit of extended radio emission with respect to all estimators of
nuclear activity compared to 3CR/FR~I.

It must also be stressed that we always performed comparisons based on the
total radio luminosity. For many CoreG, particularly those with the highest
core dominance, the emission in the radio band is dominated by the radio core
even at low frequencies, see for example the case of UGC~7386. 
Therefore, the derived deficits of extended radio emission should be 
considered as lower limits. 

\section{Revealing the bulk of the radio-loud AGN population}
\label{bulk}

\begin{figure*}
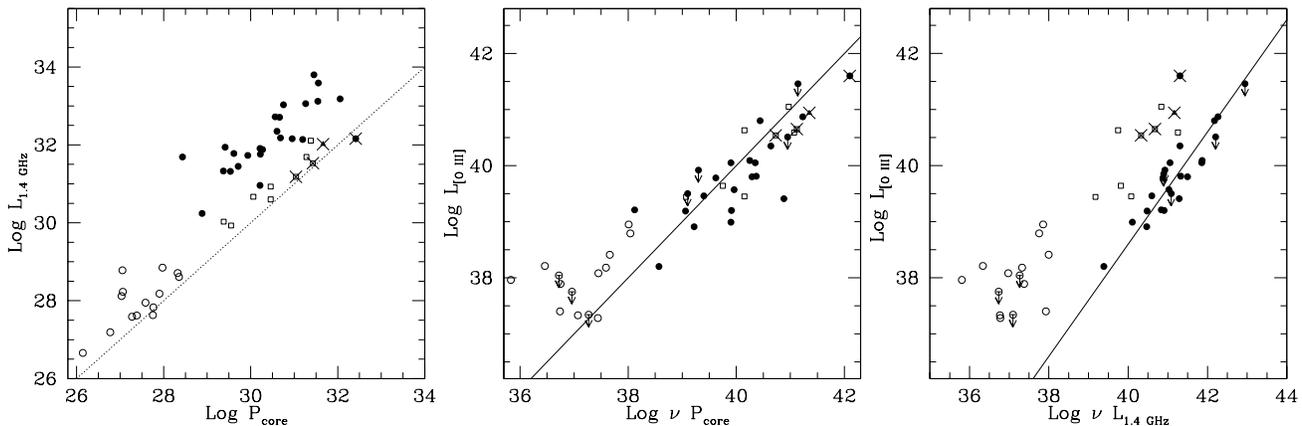

\centerline{\includegraphics[scale=0.29]{13021f5a.epsi}
\includegraphics[scale=0.29]{13021f5b.epsi}
\includegraphics[scale=0.29]{13021f5c.epsi}}
\caption{Comparison of [O III] line, core, and total radio luminosities for
  CoreG (empty circles), {\bf 3CR/FR~I} (filled circles), and a sub-sample of high core
  dominance B2 radio galaxies (empty squares). The crossed symbols locate the
  BL Lac objects in the B2 (B2~1101+38 and B2~1652+39A) and 3CR samples
  (3C~371, marked with a triangle, and 3C~84). The lines represent a constant
  ratio between the two quantities; in the left panel the dotted line marks
  the locus of equal core and total radio emission.}
\label{b2fig}
\end{figure*}

\begin{figure*}
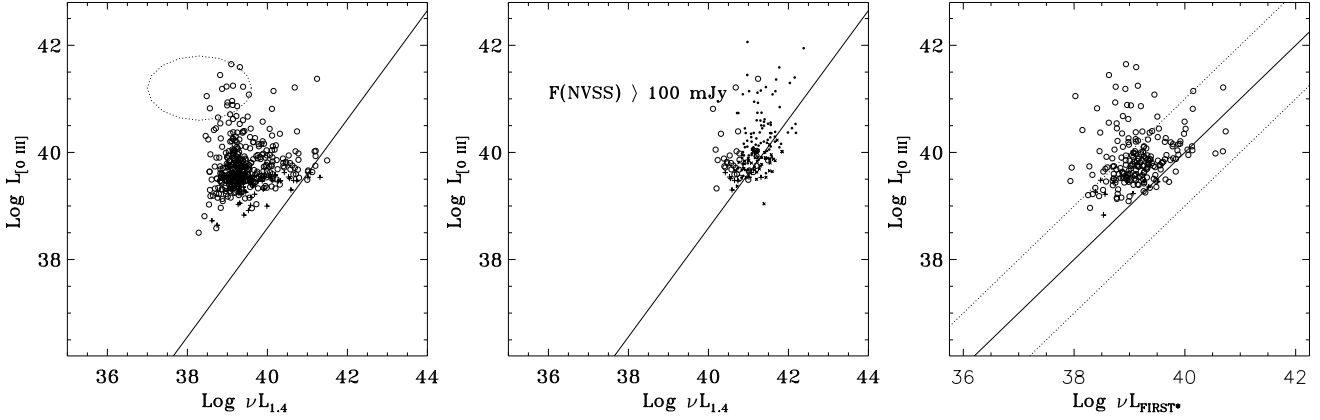

\centerline{\includegraphics[scale=0.315,angle=90]{13021f6a.epsi}
\includegraphics[scale=0.315,angle=90]{13021f6b.epsi}
\includegraphics[scale=0.315,angle=90]{13021f6c.epsi}}
\caption{Left panel: comparison of [O III] luminosity with total radio power
  for the SDSS/NVSS AGN sample (from \citealt{best05a}) with 0.03$<$z$<$0.1.
  The plus symbols are upper limits in [O III].  The solid line marks the
  equal ratio between the two quantities normalized to the 3CR sources.  The
  ellipse represents the location of Seyfert galaxies \citep{nagar99} in this
  plane. Middle panel: same as the left panel but restricted to sources with
  $F_{1.4}>$ 100 mJy, equivalent to the selection threshold of the B2 sample.
  Higher redshift sources (0.03$<$z$<$0.1) are also included as dots (or
  crosses for [O III] upper limits). Right panel: luminosity of the central
  component in the FIRST images, $L_{\rm FIRST*}$, as a proxy for the radio
  core (we considered only objects in which the size of this component is
  smaller than 3\arcsec), versus line luminosity. The dotted lines in the
  right panel indicate a change of the ratio by $\pm$ 1 dex. }
\label{best}
\end{figure*}

There are several possible alternatives to account for the different
fractional level of extended radio emission with respect to the radio core
observed in CoreG.

The first relies on the fact that CoreG have fainter radio cores than FR~I.
It can be envisaged that the ability of a RLAGN to produce a large scale
jet is suddenly reduced below a critical nuclear luminosity, as suggested by
the sharp drop in L$_{1.4}$ occurring at P$_{\rm core} \sim 10^{28.5}$ erg
s$^{-1}$ Hz$^{-1}$ (see Fig. \ref{cd}, right panel).

To test this possibility we looked for highly nucleated radio sources of
higher luminosity than the CoreG considered here. In the 3CR sample widely
used here as a benchmark, there are only two such sources, namely 3C~084 and
3C~371. However, in both cases the high $P_c$ values are most likely due to
Doppler boosting, since these galaxies host a highly polarized and variable
optical nucleus \citep{miller75,martin76}, typical of BL Lac objects.

We then considered another well studied sample, formed by the B2 radio
galaxies \citep[see][ for its definition]{fanti87}.  We selected the sources
with the highest values of $P_c$ from \citet{giovannini88} setting an
arbitrary threshold at $P_c \gtrsim 0.1$,\footnote{The core dominance
estimated from \citet{giovannini88} is referred to the total luminosity at 408
MHz.  This different definition is irrelevant for our purpose to find highly
core dominated sources.} yielding 11 objects.  B2~0648+27 was later resolved
into a compact ($\sim 1\arcsec$) steep spectrum (CSS) double
\citep{morganti03}, while B2~1144+35B is a Gigahertz Peaked Source (GPS,
\citealt{torniainen07}) with a peculiar radio-morphology \citep{giovannini07};
these objects will not be further considered here.  One of them (B2~1217+29)
is instead already included in our CoreG sample (as UGC~7386). For the
remaining 8 radio galaxies we give their multi-band luminosities in Table
\ref{b2}.

We compared the core and the total radio emission (Fig. \ref{b2fig}, left
panel). Their core luminosity extends over the range 10$^{29.5-31.5}$ erg
s$^{-1}$ Hz$^{-1}$, three orders of magnitude higher than CoreG, while
covering essentially the same range of the 3CR/FR~I sample, and, by
construction, lying close to the line of equal core and total luminosity.

From the point of view of line emission, they are located in the same region
of the L$_{\rm [O III]}$ vs P$_{\rm core}$ plane of the 3CR/FR~I sources (Fig.
\ref{b2fig}, middle panel) following the same relationship. However, they have
a deficit of total radio luminosity at a given line luminosity of an average
factor of $\sim$ 30 (Fig. \ref{b2fig}, right panel), a result similar to what
was found by \citet{morganti92}. As already mentioned above, this value should
be considered as a lower limit in terms of {\sl extended} radio emission vs
line luminosity.

These B2 radio galaxies represent the high luminosity counterparts of CoreG
from the point of view of their multi-wavelength properties, thus contrasting
with the proposed scenario that only AGN of low core power produce sources
with feeble extended radio emission.

The second possibility ascribes the presence of AGN with high and low $L_{\rm
  [O III]}/L_{\rm ext}$ ratios to the existence of two distinct populations.
  They could arise from, for example, a different mechanism of jet launching
  or of jet composition in the two classes. This predicts a bimodal
  distribution of core dominance within a given sample. Such an effect is not
  observed either in the B2 or in the 3CR samples \citep{feretti84}.

Our preferred interpretation is that, instead, the prevalence of low core
dominance sources in low frequency flux limited samples is due to a selection
bias. For example, a radio source is part of the 3CR when its total flux
exceeds $>$ 9 Jy \citep{spinrad85}. The sole core flux of 3CR sources never
reaches this threshold (with the only exception of 3C~084 with $F_{\rm core}
\sim 42$ Jy). Thus, a significant contribution from the extended emission is
required to be part of the 3CR, favoring AGN with low $L_{\rm [O III]}/L_{\rm
ext}$ ratios. The lower flux threshold of the B2 (250 mJy at 408 MHz,
\citealt{fanti78}) allows the inclusion of a larger fraction of core dominated
galaxies. The higher frequency and the much lower flux limit (1 mJy at 5 GHz),
imposed on our CoreG sample, drastically reduces the selection bias against
sources with high $L_{\rm [O III]}/L_{\rm ext}$ ratios. As a result, the CoreG
sample is formed predominantly by galaxies of high core dominance\footnote{ We
remind that three nearby 3CR sources (UGC~7360 alias 3C~270, UGC~7494 alias
3C~272.1, and UGC~7654 alias 3C~274) are part of the initial CoreG sample.}.

In order to test this scenario, we now include in the analysis the RLAGN
selected by \citet{best05a} cross-matching the $\sim$212000 galaxies drawn
from the SDSS-DR2 with the NVSS and FIRST\footnote{Sloan Digital Sky Survey,
  \citep{york00}, National Radio Astronomy Observatory (NRAO) Very Large Array
  (VLA) Sky Survey \citep{condon98}, and the Faint Images of the Radio Sky at
  Twenty centimeters survey \citep{becker95} respectively.} radio surveys.
Leaving aside 497 sources identified as star forming galaxies, this yields
2215 radio luminous AGN brighter than 5 mJy at 1.4 GHz. In Fig. \ref{best} we
show their location in the $L_{\rm [O III]}$ vs $L_{\rm1.4}$ plane limited to
the 425 sources with 0.03$<$ z$<$ 0.1, for consistency with \citet{best05a}.
Most sources are located around $L_{\rm [O III]} \sim 10^{39.5}$ erg s$^{-1}$,
while spanning a range in $L_{\rm 1.4}$ of 3 orders of magnitude, $L_{\rm 1.4}
\sim 10^{38.5} - 10^{41.5}$ \ergsHz. The behavior of higher redshift sources,
up to z=0.3, is similar but shifted at higher radio and [O III] power by
$\sim$ 0.7 dex. The vast majority of the objects lie a factor of $\sim 100$ to
the left of the relationship between line and radio emission defined by the
3CR sources and, actually, the source density increases at lower $L_{\rm [O
  III]}/L_{\rm 1.4}$ ratios. The bulk of the population of this SDSS/NVSS AGN
sample thus shows a large deficit of total radio emission, similar to that
observed in CoreG.

For these sources there are no measurements of the radio core fluxes. A
possible proxy for the radio core is the flux of the central component seen in
the FIRST images, $F_{\rm FIRST*}$, tabulated by \citet{best05a}. We
considered only the 256 objects in which the size of this component is smaller
than 3\arcsec, at most marginally resolved at the 5\arcsec\ resolution of the
survey. The comparison between $L_{\rm [O III]}$ and $L_{\rm FIRST*}$ is shown
in Fig. \ref{best}. With the caveat that this is not a genuine measurement of
the radio core, we note that the bulk of this sample has a ratio of the line
with respect to the central radio component that is slightly larger, by a
factor of $\sim$ 3, than those the 3CR and B2 samples discussed above, but
similar to that measured in the lower luminosity CoreG\footnote{There are a
minority of objects, $\sim$ 10\% of the sample, forming a substantial tail
toward high values of $L_{\rm [O III]}$, a point to which we will return in
the next Section.}.

A simple exercise can prove the selection biases discussed above. From the
NVSS/SDSS sample, we extracted the radio sources that would meet the radio
selection criterion imposed on the B2 sample, i.e.  $F_{408 \rm MHz} > 250$
mJy, that translates into $\sim$ 100 mJy at the 1.4 GHz NVSS, frequency having
adopted a radio spectral index of 0.7. The resulting sample, including
sources up to $z = 0.3$, is shown in the right panel of Fig.  \ref{best}.  We
selected only sources lying along the relation between $L_{1.4}$ and
$L_{\rm [O III}$ defined by the 3CR sources, while all sources with a deficit
of total radio emission are discarded.

We conclude that when the selection biases used are less severe (lower flux
threshold and/or higher frequency), core dominated radio galaxies emerge as
the dominant constituent of the population of RLAGN.  Conversely, the
most studied catalogues of radio galaxies, selected at high fluxes and low
frequencies, are composed of the minority of RLAGN that meet the
physical conditions required to form extended radio sources.

\begin{table}
\caption{Properties of the B2 sample.}
\begin{center}
\begin{tabular}{l | c c c c}
\hline
Name  & $F_{\rm 1.4 GHz}$ & $F_{\rm core}$ & $F_{\rm [O~III]}$ & $P_c$ \\
\hline	                              
   B2~0055+30    & 30.93$^{a}$  & 30.46 & 39.45$^{a}$ & 0.34 \\
   B2~0222+36    & 30.60$^{b}$  & 30.46 & 40.63$^{b}$ & 0.72 \\
   B2~0722+30    & 29.93$^{c}$  & 29.55 &  $-$        & 0.42 \\
   B2~1101+38    & 31.18$^{d}$  & 31.04 & 40.54$^{c}$ & 0.72 \\
   B2~1557+26    & 30.67$^{a}$  & 30.06 & 39.64$^{d}$ & 0.25 \\
   B2~1638+32    & 32.11$^{a}$  & 31.38 & 40.59$^{d}$ & 0.19 \\
   B2~1652+39A   & 31.53$^{d}$  & 31.43 & 40.65$^{e}$ & 0.79 \\
   B2~2116+26    & 30.03$^{d}$  & 29.38 & 39.44$^{f}$ & 0.22 \\
\hline
\end{tabular}

\medskip

Column description: (1) name; (2) total luminosity at 1.4 GHz in
  \ergsHz. Ref: (a) \citet{white92}; (b) \citet{condon98}; (c) \citet{FIRST};
  (d) \citet{condon02}; (3) core luminosity at 5 GHz in \ergsHz\ from
  \citet{giovannini88}; (4) nuclear [O~III] luminosity in erg s$^{-1}$. Ref:
  (a) \citet{ho97}; (b) Buttiglione et al. (2010); (c) [O~III] equivalent
  width from \citet{jansen00} and $\lambda$5007 continuum from
  \citet{marcha96}; (d) from SDSS spectrum; (e) from Ha+[N~II] measurement of
  \citet{marcha96} adopting [O~III]/(H$\alpha$+[N~II]) $\sim$ 0.38
  (H$\alpha$/[O~III]=1 and [N~II]/H$\alpha$ =1.6) \citep{buttiglione09b}; (f)
  from H$\alpha$+[N~II] image \citep{capetti:cccriga} with emission
  line ratios specified above; (5) core dominance, $P_c$ = $P_{\rm
    core}/L_{1.4}$.
\label{b2} 
\end{center} 
\end{table}

\section{Discussion}
\label{discussion}

The results discussed in the previous sections indicate that in RLAGN all
indicators of nuclear activity, i.e. line luminosity, power of the radio core,
and (when available) the luminosity of the optical and X-ray nucleus, are all
closely and quasi-linearly correlated to each other despite the fact that they
cover a range of $\sim$ 6 orders of magnitude. This is evidence that from the
point of view of the nuclear properties, the RLAGN population is very
homogeneous.

Conversely, RLAGN present an extremely large range of total radio power, a
factor of $\sim$ 1000, at a given level of radio core or emission line
luminosity. The vast majority of the population is associated with sources of
relatively low levels of extended radio emission with respect to classical FR~I
radio galaxies. Even the NVSS/SDSS sample is not
completely free from selection biases and that the space density of core
dominated sources might be even higher, and extending to more extreme ratios
of core versus extended emission.

The most studied catalogues of radio galaxies are severely biased against the
inclusion of objects with high core dominance, since a large contribution from
extended emission is needed to fulfill the stringent flux requirements of low
frequency, high flux threshold samples. They are thus composed of the minority
of AGN that meets the physical conditions required to form extended
radio sources, while the bulk of the RLAGN population is virtually
unexplored.

It is important to stress that these AGN, despite the lower level of extended
radio emission, are not radio-quiet. Not only do they have radio-loud nuclei,
but they have (on average) a ratio between radio and line luminosity larger by
a factor of 300 with respect to radio-quiet AGN of similar radio
luminosity. This is clearly seen in Fig. \ref{best}, where we reproduce the
location of the sample of Seyfert galaxies studied by \citet{nagar99}.

From the point of view of the host galaxies of the core dominated
radio galaxies, for the sample of Core galaxies they are massive early-type (E
and S0) galaxies, less luminous than the 3CR/FR~I host on average by 1
magnitude, but the luminosity ranges of the two classes overlap considerably
\citep{balmaverde06b}. Their black hole masses are 10$^{7.8-9.5}$ M$_{\sun}$,
with a distribution indistinguishable from those measured in 3CR/FR~I
hosts. The SDSS/NVSS sample of \citet{best05a} is also formed predominantly by
massive objects (M $\sim$ 10$^{10-12}$ M$_{\sun}$, with an average of
$\sim$10$^{11.5}$ M$_{\sun}$) and with black hole mass in the range
10$^{6.8-9.5}$ M$_{\sun}$ (average value of $\sim$10$^{8.5}$ M$_{\sun}$),
similar to those of 3CR/FR~I. A detailed study of the
morphology and colors of the SDSS/NVSS hosts will be the subject of a
forthcoming paper, but, apparently, high and low core dominance radio sources
cannot be readily separated on the basis of differences in their hosts.

The environment could also play an important role. Analyzing the literature,
  we found that CoreG typically are the brightest galaxies of groups of
  intermediate richness. With respect to 3C/FRI they seem to avoid the center
  of rich clusters. Instead, at the moment there are no indications in the
  environment of the bulk of the SDSS/NVSS radio-loud AGN population.

One possibility to account for the sequence from low to high extended power,
at the same level of nuclear activity, is to assume a relation with the source
age, in the sense that low power sources are younger that those of higher
power. This requires that the radio luminosity increases with time as the
radio source expands. Furthermore, the growth of the radio luminosity must
depend strongly on age to reproduce the relatively higher density of sources
with the lowest radio power. This requirement contrasts with what is derived
for the luminosity evolution of CSS and GPS sources (see e.g.
\citealt{snellen00}), a general result based on self-similarity
and on the formation of pressured confined radio-lobes.  Nonetheless, lower
power sources might form predominantly plume-like radio structures and all is
further complicated by the possibility of recurrent outflows, slowly burrowing
their way into the interstellar medium of the host galaxy.

In a similar line of interpretation, the high number of young/faint sources is
expected if they are short lived and they never grow to large scales.
\citet{alexander00} presented an analytical model for the evolution of radio
sources from small physical scales to classical doubles; they found that
observational data can be reproduced assuming the presence of a population of
sources that suffer disruption of their jets before escaping the host core
radius.  This scenario confirms earlier predictions that compact symmetric
objects (CSOs), with typical sizes of $\sim$ 100, could switch off after a
short period of time, $\sim 3000$ years, possibly due to a lack of sufficient
fueling \citep{readhead94,kunert09}.  Similarly, \citet{reynolds97} proposed a
model in which radio sources are intermittent on timescales of $\sim
10^{4}-10^{5}$ years.  If this is the correct interpretation, the question is
why classical FR~I are active over large timescales, necessary to form radio
structures up to the Mpc scale, while the majority of the RLAGN population is
short lived.

The very large range of total radio-power at a given level of emission line
luminosity lead \citet{best05b} to argue that they are independent phenomena,
triggered by different physical mechanisms. Indeed our results point to the
conclusion that the total radio emission is not simply determined by the AGN
activity level. Conversely, the radio-core luminosity, despite the effects of
beaming, is closely linked to line emission and it is a rather good indicator
of nuclear activity. \citet{capetti:cccriga} argued that the strong
correlation between line and optical continuum nuclear emission found for
3CR/FR~I radio galaxies suggests that the optical cores (most likely of
non-thermal origin) can be directly associated with the source of ionizing
photons, i.e.  that we are seeing a jet-ionized narrow line region. This
suggests that the primary link between radio core and line luminosity is due
to the relationship existing between both quantities with the optical core.
Conversely, the well established correlation between total radio and line
luminosities \citep[e.g.][]{baum89b,rawlings89,rawlings91} is likely to be the
result of the selection of radio sources of similar range of core dominance,
imposed by the selection criteria of the sample.

The indication that the bulk of the RLAGN population is not represented in the
well studied samples of radio galaxies also requires us to revise profoundly
the unified models based on orientation for the low luminosity RLAGN (see e.g.
\citealt{urry95}).  The behavior of the two 3CR sources with the highest core
dominance (3C~084 and 3C~371) is very instructive in this context. They show
an excess of the radio core with respect to the sources of similar $L_{\rm [O
III]}$, as expected considering the Doppler boosting of their nuclear
emission.  Nonetheless, they have a strong deficit in total radio emission,
given their [O III] luminosity, despite the enhanced core contribution.  The
same result applies to the two well known BL Lac objects that is part of the
B2 sample (B2~1101+38 alias MRK~421 and B2~1652+39A alias MRK~501). The
beaming effects favor their inclusion in the flux limited samples but these
sources, if they were observed along a line of sight forming a larger angle
with the jet axis, would show an even larger offset from the relationship
between radio and line luminosity defined by the 3CR sample. The small number
statistics suggests some caution, but this supports the idea that the parent
population of BL Lac sources, and thus the overall population of radio
galaxies, is dominated by sources with high $L_{\rm [O III]}/L_{\rm ext}$
ratios.  Furthermore, our results indicate that a high core dominance in a
radio source cannot be taken as sole evidence to deduce a strong Doppler
boosting, in line with the suggestion of \citet{marcha05}.

The vast population of core dominated radio galaxies might also have
consequences on the predictions and interpretation of
the density of high radio frequency extragalactic sources 
(see e.g. \citealt{dezotti05}) relevant for the experiments on the Cosmic
Microwave Background, and also on the association between AGN and the
high energy cosmic rays (see e.g. \citealt{abraham07}).

We conclude this section considering the minority of objects in the SDSS/NVSS
sample form a substantial tail toward high values of $L_{\rm [O III]}$ with
respect to the power of the central FIRST component (see Fig. \ref{best},
right panel), our proxy for the core power. These sources are located in the
region where radio-quiet AGN are found (see \citealt{capetti06,capetti07}),
being characterized by larger $L_{\rm [O III]}/P_{\rm core}$ ratios than radio
loud AGN. They have radio luminosities $L_{1.4} \sim 10^{29.5-31}$ erg
s$^{-1}$ Hz$^{-1}$, well within the range of Seyfert galaxies
\citep{ulvestad84}, suggesting a possible contamination of the SDSS/NVSS
sample by radio-quiet AGN. Clearly these objects deserve further study,
particularly concerning their optical spectroscopic classification and their
radio structure, aimed to accurate radio core measurements.

\section{Summary and conclusions}
\label{summary}

We considered a sample of 14 luminous and nearby early-type galaxies hosting
miniature RLAGN of extremely low radio luminosity (10$^{27-29}$ erg s$^{-1}$
Hz$^{-1}$ at 1.4 GHz) for which an extensive radio and optical analysis is
performed. We collect the multi-wavelength nuclear information for our sample of core
galaxies in order to compare them with those of the more powerful FR~I
radio galaxies.

A radio analysis of these sources reveals that in many CoreG, the extended
radio morphology is indicative of a collimated outflow. However, they are
substantially less extended (up to $\sim$ 20 kpc) and their core dominance is
a factor of 40-100 higher than in more powerful FR~I radio sources in the 3CR
sample. Their radio spectral properties are instead similar to those of 3CR/FR~I.

We also obtained new optical spectroscopic observations. 
CoreG show emission line ratios typical of AGN (they can be classified as low
excitation galaxies), similar to, but of even lower excitation than, those of
3CR/FR~I radio galaxies, and matching more closely those of radio-quiet AGN of
the same of line luminosity. This indicates that the line ratios are driven
mostly by the AGN luminosity rather than by its radio-loudness.

The main result obtained from the spectroscopic study is that the observed
line emission has generally an AGN origin and this enables us to include these
emission lines in our analysis of the AGN properties. While miniature radio
galaxies follow similar relationships to those found for more powerful radio
galaxies in terms of line, optical, and radio nuclear luminosities, as well as
accretion rate, they have a deficit of a factor of $\sim$100 in extended
radio emission with respect to that of classical 3CR/FR~I.  From the radio
data alone, the possibility that this is due to a Doppler boosting enhancement
of the radio core emission was still viable. The inclusion of emission lines
(independent of orientation) leads us to conclude that what we are seeing in
CoreG is a genuine deficit of extended radio emission, considering different
estimators of nuclear activity, with respect to 3CR/FR~I.

The deficit of extended radio emission is also found in radio sources
extracted from the B2 sample, with radio cores as powerful as those of
classical FR~I radio galaxies. Therefore, the difficulty to produce prominent
extended radio structures is not simply due to a low AGN luminosity. Actually,
core dominated sources form the bulk of radio-loud AGN population in the
SDSS/NVSS sample. The vast majority of these sources show, at a given level of
line luminosity, a deficit of radio emission of a factor of $\sim$ 100.
Furthermore, this value should be considered as an upper limit due to the
significant contribution of the radio core and since the samples considered
are not completely free from selection biases.

At a given level of nuclear emission, one can find radio sources with an
extremely wide range of radio power. Nonetheless, even the objects with the
lowest level of extended radio emission are not radio-quiet. Not only are
their nuclei radio-loud, but they have on average a ratio of radio to line
luminosity larger by a factor of 300 with respect to radio-quiet AGN of
similar radio power.

One possibility to explain this effect is to assume that this is driven by the
source age, assuming that the radio luminosity increases with time as the
radio source expands. A high number of young/faint sources is expected if the
vast majority of RLAGN is short lived and never grows to large scales.  If
this is the correct interpretation, the questions are: what mechanism sets the
lifetime of the radio galaxies? Why are most radio-loud galaxies short-lived,
while only a minority is active over long timescales?

The prevalence of low core dominance sources in flux-limited samples is
apparently due to a selection bias, since the inclusion of sources with
luminous extended radio structures is highly favored.  This result has several
important ramifications. For example, unified models for RLAGN should be at
least in part reconsidered, since the bulk of the RLAGN population is not well
represented in the most studied samples of radio galaxies. Similarly, the
relationship between line and radio emission results at least in part from
selection effects, while apparently the link between the line and radio core
is more robust and relies on the common dependence of these quantities on the
strength of the nuclear continuum emission.

A thorough study of the morphology and colors of the hosts of core dominated
sources, in particular those that are part of the large SDSS/NVSS sample, as
well as of their radio morphology and optical spectroscopic classification is
clearly required. This will enable us to perform a more detailed comparison
with classical extended radio galaxies to reveal further similarities or
differences that might account for their diversity.

The samples considered here are limited to AGN with $L_{\rm [O III]} \lesssim
10^{41}$ erg s$^{-1}$. It would be of great interest to explore
whether the dominance of radio galaxies with relatively low extended radio
emission applies also to sources of higher power.

\appendix
\label{append}
\section{Spectroscopic observations of core galaxies}

\begin{figure*}
\includegraphics[scale=0.90]{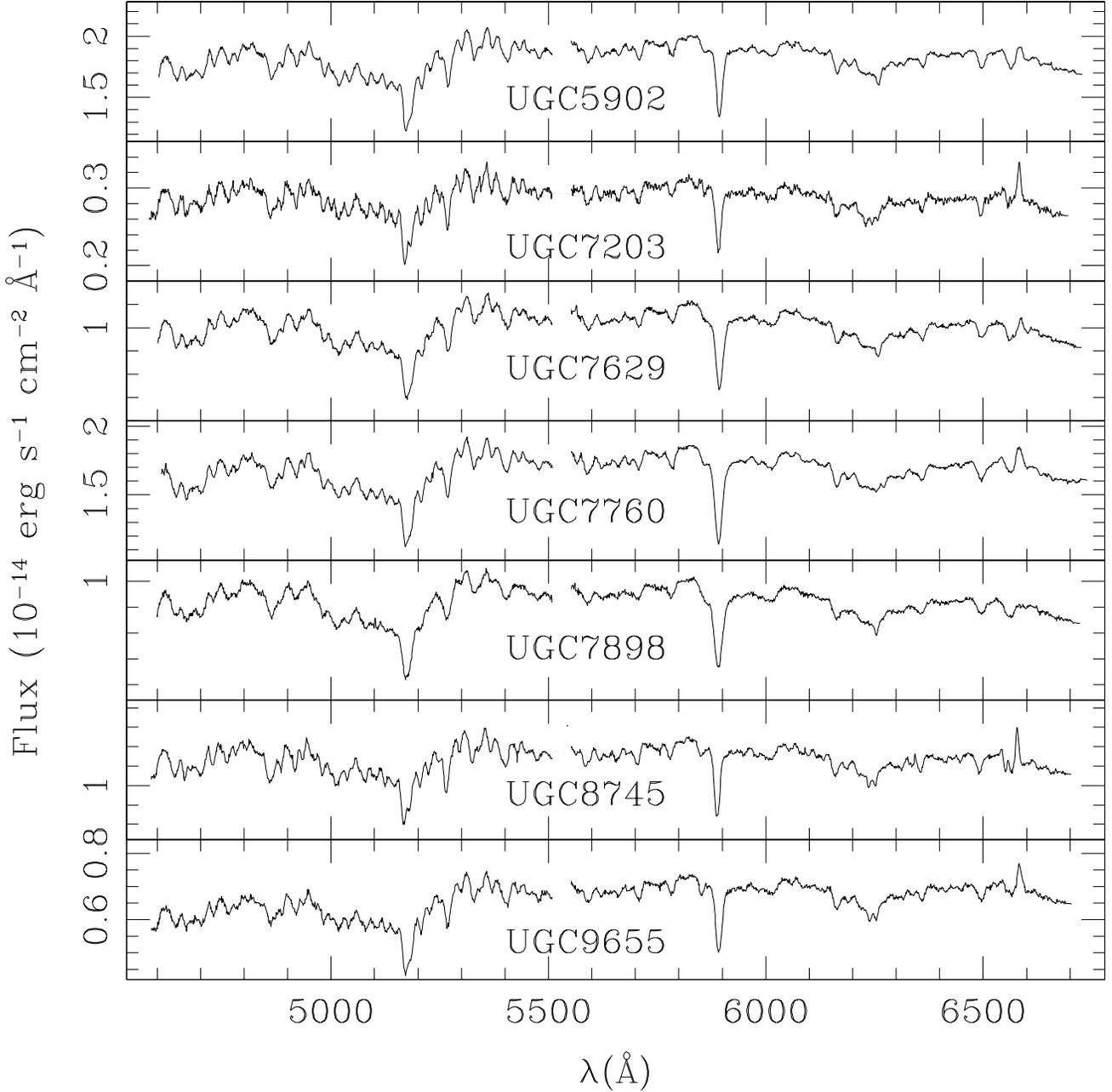}
\caption{Spectra of the 7 sources observed at the TNG. The wavelength scale
  has been corrected for the redshift reported in Table \ref{tab1}. For each
  object the flux scale covers the range (0.6 - 1.2) $\times
  F_{\lambda=5500}$.}
\label{spettri}
\end{figure*}

On 2006 10th March we observed a subsample of 7 CoreG with Telescopio
Nazionale Galileo (TNG), a 3.58 m optical/infrared telescope located on the
Roque de los Muchachos in La Palma, Canary Islands (Spain).  We selected the
sources for which the spectra available from the survey of \citet{ho97} are of
insufficient quality to extract accurate line measurements and to obtain a
robust spectral classification. The available spectrum of an additional galaxy,
UGC~968, is also of rather poor quality, but was not visible during the
observing run. 

The observations were made using the DOLORES
(Device Optimized for the LOw RESolution) spectrograph installed at the
Nasmyth B focus of the telescope, with an exposure time of 1200 s. The chosen
long-slit is 2\arcsec wide. For each target we used the VHR-V grism which has
a dispersion of 1.05 \AA/pixel, a spectral coverage of 4650-6800 \AA\ and a
resolution of $R\sim$800.

The spectra were reduced using the LONGSLIT package of
IRAF\footnote{IRAF is the Image Reduction and Analysis Facility of the
National Optical Astronomy Observatories, which are operated by AURA, Inc.,
under contract with the U.S. National Science Foundation. It is also available
at http://iraf.noao.edu/.}. The optical spectra were processed by the standard
spectroscopic calibration (bias subtraction, flat field normalization,
background subtraction, wavelength calibration, and flux calibration with
spectrophotometric standard stars). We then summed and extracted a region of
2\arcsec along the spatial direction, resulting in a region covered by our
spectra of 2\arcsec$\times$2\arcsec.

The spectra obtained with this aperture contain emission from the active
nucleus as well as a substantial contribution from the host galaxy stellar
population. To proceed in our analysis it is necessary to separate
these two components by subtracting the starlight from the 
extraction aperture. 

\begin{figure}
\includegraphics[scale=0.45]{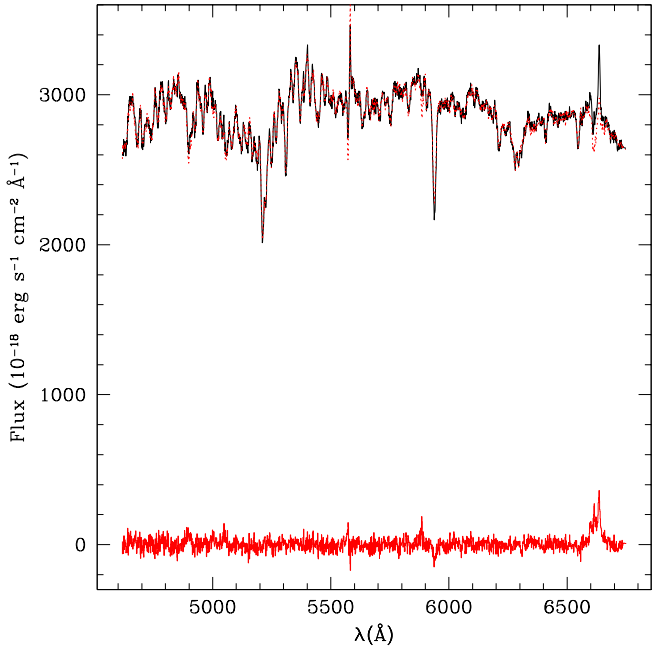}
\caption{Spectrum of UGC~7203 as an example of the subtraction of galaxy
  emission (red dashed line) from the nuclear spectrum (black solid
  line). The residuals, shown on the bottom, provide the genuine spectrum of
  the AGN.}
\label{7203spettro}
\end{figure}

In order to estimate the contribution of stellar emission, we used as a
template the spectra extracted from two off-nuclear regions, 2\arcsec\ wide,
flanking the nuclear aperture. We then appropriately scaled the template to
match the nuclear spectrum, by excluding from the match the spectral regions
corresponding to emission lines, as well as other regions affected by telluric
absorption, cosmic rays or other impurities. The subtraction of this stellar
emission template from the nuclear aperture gives us the genuine nuclear
emission spectrum. This method is illustrated in Fig.~\ref{7203spettro}, where
we show an example of starlight subtraction.  Note that, since the AGN optical
core of these radio sources \citep{balmaverde06b} contributes to $\lesssim$
1/100 of the light within the extraction region, we can neglect the presence
of nuclear continuum emission in the spectra. The accuracy of the starlight
subtraction can be assessed by the absence of large-scale patterns in the
residual spectra at a typical amplitude of $<$0.5 $\times$ 10$^{-16}$ erg
s$^{-1}$ cm$^{-2}$ \AA$^{-1}$.  The noise rms is of $<$10$^{-16}$ erg s$^{-1}$
cm$^{-2}$ \AA$^{-1}$.

The next step of our analysis consists of the measurement of the emission line
intensities, for which we used the SPECFIT package in IRAF. We measured line
intensities fitting Gaussian profiles to H$\beta\lambda$4861,
[O~III]$\lambda\lambda$4959,5007, [O~I]$\lambda\lambda$6300,6364,
H$\alpha\lambda$6563, [N~II]$\lambda\lambda$6548,6584 (Table~\ref{tabriga}).
Some constraints were adopted to reduce the number of free parameters: we
required the width5 and the velocity to be the same for all the lines. The
integrated fluxes of each line were free to vary except for those with known
ratios from atomic physics: i.e. the [O~I]$\lambda\lambda$6300,64,
[O~III]$\lambda\lambda$4959,5007 and [N~II]$\lambda\lambda$6548,84 doublets.
After subtraction of the narrow line components we did not find significant
residuals around the Balmer lines that might indicate the presence of a broad
line region.  In Table~\ref{tabriga} we report the line measurements for these
7 objects as well as those presented by \citet{ho97} for the remaining sample.

\begin{table*}
\caption{Optical emission line fluxes of the core galaxies.}
\begin{center}
\begin{tabular}{c | c c c c c c c}
\hline
Name     & V    & H$\beta$  & [O III] & [O I] & H$\alpha$ & [N II] & [S~II] \\
\hline
UGC~5902 & 950  &3.2$\pm$0.5  & 4.8$\pm$0.4 & $<$1.2      & 6.2$\pm$0.7   & 3.8$\pm$0.4 & $-$ \\
UGC~7203 & 2379 &0.9$\pm$0.1  & 1.0$\pm$0.1 & 0.5$\pm$0.1 & 2.8$\pm$0.1   & 4.4$\pm$0.2 & $-$ \\
UGC~7629 & 1047 &0.9$\pm$0.2  & 1.1$\pm$0.2 & $<$0.6     & 1.6$\pm$0.2   & 3.3$\pm$0.2 & $-$\\
UGC~7760 & 450  &2.0$\pm$0.5  & 4.4$\pm$0.7 & $<$1.2      & 10$\pm$1      & 20.0$\pm$0.4& $-$ \\ 
UGC~7898 & 1198 &0.8$\pm$0.2  & $<$0.7     & $<$0.8     & 0.6$\pm$0.2   & $<$1.2      & $-$ \\    
UGC~8745 & 2054 &3.0$\pm$0.3  & 2.8$\pm$0.3 & $<$1.1      & 6.4$\pm$0.3   &12.0$\pm$0.3 & $-$ \\
UGC~9655 & 2037 &1.1$\pm$0.3  & $<$1.2      & $<$0.91     &1.6$\pm$0.3    & 3.3$\pm$0.3 & $-$ \\ 
\hline	       									
UGC~0968  & 2422 & 0.94$^{b}$&  $<$0.45 & $<$0.42  &  3.0 & 4.7   &  2.1  \\     
UGC~6297 & 1025 & 3.2       &   7.3    &   3.4    &  18  & 38    &  15 \\
UGC~7386 & 788  & 51        &  67      & 54       &   132&  161  &  195  \\
UGC~7797 & 2283 & 2.4       &    5.6   &  4.6     &   11 & 16    &  19   \\
UGC~7878 & 1012 & 1.9       &   3.6    &  1.2     &  5.4 &  9.9  &  7.8  \\       
UGC~9706 & 1832 & 1.7       &   2.1    & 0.6$^{c}$&  6.6 & 10    &  6.3 \\     
UGC~9723 & 950  & 0.54      &  1.1     &  $<$0.47 &  3.6 &  4.9  &  2.7 \\
\hline
\end{tabular}

\medskip

Column desctription: (1) name; (2) radial
  velocity (cz) corrected for Local Group infall onto Virgo cluster [km
  s$^{-1}$]; emission line flux in units of 10$^{-15}$ erg s$^{-1}$ cm$^{-2}$:
  (3) H$\beta\lambda$4861, (4) [O~III]$\lambda$5007, (5) [O~I]$\lambda$6300,
  (6) H$\alpha\lambda$6563, (7) [N~II]$\lambda$6584, (8) sum of
  [S~II]$\lambda$6716 and [S~II]$\lambda$6731. The upper limit fluxes are
  measured at 3$\sigma$.  The data below the horizontal line are from
  \citet{ho97}. For these sources, values with uncertainties of $\pm$30\%-50\%
  are marked by `b', while highly uncertain values with probable errors of
  $\pm$100\% are marked by `c'.

\label{tabriga}
\end{center}
\end{table*}

\section{The multiphase inter-stellar medium of early-type galaxies}

As discussed in the Introduction, the level of activity in CoreG and FR~I is
closely related to the accretion rate of hot gas, that represents the dominant
process in powering these radio sources.  The line emission detected in the
sources of our sample can be used to estimate the amount of relatively cold
ionized gas present in their nuclear regions, to be then compared with the hot
ISM component.

Following \citet{osterbrock89}, in the case B approximation, we have:
$$
M_{\rm H~II} = \frac{m_{p}+0.1 m_{He}}{n_{e} \alpha^{eff}_{H\beta} h\nu_{H\beta}} L_{H\beta} \approx \frac{3 \times 10^{-36}}{n_{3}} L_{H\alpha} \,\,M_{\sun}
$$
where $n_3$ is the gas electronic density in 10$^3$ cm$^{-3}$ units
and we assumed a recombination
coefficient of $\alpha^{eff}_{H\beta}$ = 3.03 $\times$ 10$^{-14}$ cm$^{3}$
s$^{-1}$, i.e. a temperature of 10$^4$ K, and $L_{H\beta} = L_{H\alpha}$/3.
The resulting masses are in the range 40 - 1800 $M_{\sun}$.

On the other hand, we can use the results of \citet{balmaverde08} to estimate
the mass of the X-ray emitting hot gas in the galactic coronae. From the
brightness profiles of the Chandra images, they derived the gas density
profiles that can be integrated out to the same physical radius $r_{\rm spec}$
used to extract the optical spectra,
$$
M_{\rm hot} = 2 \pi n_{B} r^{3}_{B} \left( \frac{r_{\rm spec}}{r_{B}}\right)^{2} 
$$ where $n_{B}$ is the electronic density at the Bondi radius $r_{B}$, having
used the averaged logarithmic slope ($\alpha \sim -1$) of the density profiles
to perform the integration.

The masses of the hot gas are in the range 2$\times 10 ^4 - 5 \times 10^6$
M$_{\sun}$, while the ratios between ionized and hot gas range from $10 ^{-4}$
to 7$\times 10 ^{-3}$. This result indicates that although in the ISM of these
early-type galaxies a substantial amount of ionized gas is present, the hot
ISM phase is by far its dominant component. This strengthens a posteriori the
validity of the estimates of accretion rates derived considering only the high
temperature gas.

\end{document}